\documentclass[fleqn,usenatbib]{mnras}

\usepackage{newtxtext,newtxmath}

\usepackage[T1]{fontenc}
\usepackage{ae,aecompl}
\usepackage{xcolor}

\usepackage{graphicx}	
\usepackage{amsmath}	
\usepackage[export]{adjustbox}

\newcommand{\sepspace}{\vspace*{0.9em}}		

\newcommand{\Sec}[1]{Section~\ref{#1}}

\newcommand{\hGpc}{\,h^{-1}\,{\rm Gpc}}

\newcommand{\Mpc}{\,{\rm Mpc}}
\newcommand{\hMpc}{{\ifmmode{\,h^{-1}{\rm Mpc}}\else{$h^{-1}$Mpc}\fi}}
\newcommand{\hkpc}{{\ifmmode{\,h^{-1}{\rm kpc}}\else{$h^{-1}$kpc}\fi}}
\newcommand{\hMsun}{{\ifmmode{\,h^{-1}{\rm {M_{\odot}}}}\else{$h^{-1}{\rm{M_{\odot}}}$}\fi}}
\newcommand{\Msun}{\,\rm {M_{\odot}}}
\newcommand{\Msunyr}{\,\rm {M_{\odot}}\,yr^{-1}}
\newcommand{\Mstar}{{\ifmmode{\,M_{*}}\else{$M_{*}$}\fi}}
\newcommand{\Mhalo}{{\ifmmode{\,M_{\rm halo}}\else{$M_{\rm halo}$}\fi}}
\newcommand{\ltsima}{$\; \buildrel < \over \sim \;$}
\newcommand{\gtsima}{$\; \buildrel > \over \sim \;$}
\newcommand{\lsim}{\lower.5ex\hbox{\ltsima}}
\newcommand{\gsim}{\lower.5ex\hbox{\gtsima}}

\newcommand{\rockstar}{\textsc{Rockstar}}

\newcommand{\galacticus}{\textsc{Galacticus}}
\newcommand{\sag}{\textsc{SAG}}
\newcommand{\sage}{\textsc{SAGE}}

\defcitealias{jiang}{J18}

\title[Protoclusters at $z=5.7$]{Protoclusters at $\bf z=5.7$: A view from the MultiDark galaxies}

\author[W. Cui et al.]{%
Weiguang Cui,$^{1}$\thanks{E-mail: weiguang.cui@ed.ac.uk}
Jiaqi Qiao,$^{2}$\thanks{E-mail: j.qiao@sms.ed.ac.uk}
Romeel Dav\'e,$^{1}$
Alexander Knebe,$^{3,4,5}$
John A. Peacock,$^{1}$
\newauthor and Gustavo Yepes$^{3,4}$
\\
$^{1}$Institute for Astronomy, University of Edinburgh, Royal Observatory, Edinburgh EH9 3HJ, United Kingdom\\
$^{2}$School of Physics and Astronomy, University of Edinburgh, Edinburgh EH9 3FD, United Kingdom\\
$^3$Departamento de F\'isica Te\'{o}rica, M\'{o}dulo 8, Facultad de Ciencias, Universidad Aut\'{o}noma de Madrid, 28049 Madrid, Spain\\
$^4$Centro de Investigaci\'{o}n Avanzada en F\'{\i}sica Fundamental (CIAFF), Universidad Aut\'{o}noma de Madrid, 28049 Madrid, Spain \\
$^5$International Centre for Radio Astronomy Research, The University of Western Australia, 35 Stirling Highway, Crawley, Western Australia 6009, Australia\\
}

\date{Accepted XXX. Received YYY; in original form ZZZ}

\pubyear{2020}

\begin{document}
\label{firstpage}
\pagerange{\pageref{firstpage}--\pageref{lastpage}}
\maketitle

\begin{abstract}
Protoclusters, which will yield galaxy clusters at lower redshift, can provide valuable information on the formation of galaxy clusters. However, identifying progenitors of galaxy clusters in observations is not an easy task, especially at high redshift. Different priors have been used to estimate the overdense regions that are thought to mark the locations of protoclusters. In this paper, we use mimicked Ly$\alpha$-emitting galaxies at $z=5.7$ to identify protoclusters in the MultiDark galaxies, which are populated by applying three different semi-analytic models to the $1\hGpc$ MultiDark Planck2 simulation. To compare with observational results, we extend the criterion 1 (a Ly$\alpha$ luminosity limited sample), to criterion 2 (a match to the observed mean galaxy number density). To further statistically study the finding efficiency of this method, we enlarge the identified protocluster sample (criterion 3) to about 3500 at $z=5.7$ and study their final mass distribution. The number of overdense regions and their selection probability depends on the semi-analytic models and strongly on the three selection criteria (partly by design). The protoclusters identified with criterion 1 are associated with a typical final cluster mass of $2.82\pm0.92 \times 10^{15} \Msun$ which is in agreement with the prediction (within $\pm 1 \sigma$) of an observed massive protocluster at $z=5.7$. Identifying more protoclusters allows us to investigate the efficiency of this method, which is more suitable for identifying the most massive clusters: completeness ($\mathbb{C}$) drops rapidly with decreasing halo mass. We further find that it is hard to have a high purity ($\mathbb{P}$) and completeness simultaneously.
\end{abstract}




\section{Introduction}
The Lambda Cold Dark Matter ($\Lambda$CDM) model describes the universe as a hierarchical system with smaller haloes merging to form larger ones. Galaxy clusters, as the largest gravitationally bound objects in the universe, provide ideal laboratories for cosmological studies, especially galaxy evolution studies \citep{Kravtsov2012}. At high redshift, $z$, most of the clusters have not yet formed and we thus see only their progenitors, protoclusters. Through these structures, we can have a glimpse of how matter rapidly assembles and forms present-day massive galaxy clusters \citep{overdensemethod}. 

Local galaxy clusters are mostly found by observational methods such as X-ray emission \citep{xraytheo,xrayexp} or red sequence selection \citep{redsequence}. Protoclusters, due to their lack of such clear observational signatures at high redshift \citep[for example,][]{Sembolini2014}, are mainly found by looking for areas with high galaxy density \citep{overdensemethod}. To seek these overdense regions, the general method is to survey over an area of sky and then make a cut on the data to see if the remaining luminous objects are dense enough \citep{z3photometric, 1604.08627, 1708.00447, 1804.09231}.  
Thanks to the enhancement of telescope technology, a growing number of protoclusters have been found in recent years. One class of search method relies on finding objects by chance in a general survey of a large area of sky; the other method is to target a region that is known to contain previously identified galaxies as a good tracer for the protoclusters \citep{overdensemethod}. 

As galaxy clusters are usually found at $z<2$ \citep{cluster2-1, cluster2-2, cluster2-3}, protoclusters are normally discovered over a broad range of $z \simeq 2-6$ with one of the largest samples of currently known protoclusters being concentrated around $z \lesssim 4$ \citep{1405.2620, 1604.08627, 1804.09231, 10.1093}. To identify protoclusters at higher redshifts would be much more challenging due to their low number density and the faintness of distant galaxies \citep{def}. However, simulations that can track the progenitors of the current large-mass galaxy clusters predict that galaxy protoclusters may be found at high redshift, $z\simeq6$ \citep[for example][]{simprotoat6-1, simprotoat6-3, simprotoat6-2}. On the observational side, more protoclusters have been traced toward high redshift \citep[for example][]{Venemans2002,Ouchi_2005, Toshikawa2012, Toshikawa2014, Chanchaiworawit2017, 1902.09555, Higuchi2019}. These studies have identified a large number of Ly$\alpha$-emitting galaxies (LAEs) with significantly high star-formation rates (SFR), indicating a cluster in the bursting formation phase.

Recently, using luminous LAEs, \citet[hereafter J18]{jiang} announced the discovery of an overdense region at $z=5.7$ in a spectroscopic survey of galaxies in four square degrees. 23 LAEs are located in a near-cubic volume of ${35}^3 \Mpc^3$, which is embedded in an even larger overdense region with another 41 photometric identified LAEs. This region is first observed by \cite{Ouchi_2005}, who identified it as two dense clumps of LAEs with a sphere of 1 Mpc diameter. They estimated a large overdensity, $\sim 80$, of the LASEs in this region and predicted it as a progenitor of a massive galaxy cluster. The LAEs are selected based on the narrow-band excess colour cut ($i' - NB816 > 1.0$) with an additional requirement that none should be detected in any bands bluer than $R$. The LAE number density in this ${35}^3 \Mpc^3$ volume is 6.6 times the average number density at $z=5.7$. They estimated that this region will collapse into a galaxy cluster with a mass of $3.6\pm0.9 \times 10^{15} \Msun$, making it the most massive protocluster at $z>4$ and the first of such a large structure reported at $z>5$. This discovery motivates a hunt for protoclusters at even larger redshifts, the first several billion years after the big bang. Therefore, a question is raised of whether it is possible to find other such large protoclusters at $z>5$ and if so, how common they are. 

A search for protoclusters at high redshifts observationally would constitute the proverbial needle in a haystack, requiring enormous large-scale surveys covering hundreds of square degrees. Besides, some drawbacks may also slightly weaken the effectiveness of observational searches. \cite{def} pointed out that many observations are carried out in a small sky coverage of only a few arcmins on a side owing to the restricted field of view of telescopes, which may not cover the whole cluster region and lead to an underestimate of the size, mass and overdensity of observed object. In addition, this also brings a problem concerning the measurement of the field sample. Due to the complex structure of protoclusters at high redshift, they could span a large extent in some directions; thus, a narrow window may not be able to provide a clean field sample distant enough from the central galaxy \citep{1210.8407, haines}. When it comes to the contamination from foreground and background galaxies, the most effective way today is to use the narrow-band (or Ly$\alpha$) technique to look for the wavelength of specific emission lines from the protocluster \cite[for example][]{narrowband}. However, this method also has its weaknesses, as only galaxies with high Ly$\alpha$ emission are being selected \citep{1203.2196, def, Venemans07}. This provides a biased sample for estimating densities, which may overlook a large number of galaxies in the overdense area. There is a further bias not just from the incompleteness of the tracers, but also from the method itself. As pointed out by \cite{1310.2938}, the overdensity of the protocluster would be decreased with the increase of the width of the narrow-band redshift range $\Delta z$ since the scatter in each bin lessens. Therefore, some protoclusters with insufficiently large densities could escape the attention of researchers. What aggravates the problem is that, even if we could measure an accurate overdensity, we still need to rectify the redshift-space distortions due to the projection onto a two-dimensional surface \citep{steidel}. 

With the possible inconvenience outlined previously, a safeguard would be to have an approach based on simulations. With the increase of computational ability, the resolution and volume of simulations both continue to improve, revealing a significantly larger number of simulated protoclusters for future researchers. In this paper, we will present the exploration of protoclusters at $z=5.7$ in a set of mock galaxies that are populated by three different semi-analytical models on a dark-matter-only cosmological simulation. The purpose of this is that we can easily identify the effects of different models on the number of protoclusters being found. We hope that with the three models, we can provide a robust answer on how common protoclusters are. Furthermore, tracking these protocluster regions down to $z=0$, we can predict their final states -- do they all result in such massive clusters?

This paper is organised as follows: in \Sec{sec2} we describe the catalogue used in this study and how we identify protoclusters. Using the protocluster catalogues we created, in \Sec{Result}, we make a comparison of findings between the different models and the different criteria we imposed. Following from the results, we discuss our findings and the implications for future research in \Sec{Conclusion}.

\section{Simulations and methods} \label{sec2}

\subsection{MDPL2 Simulation and SAM Catalogues} \label{Simulation}


Although hydrodynamic simulations directly provide huge amounts of valuable data, they normally require substantial computational resources -- even though the simulation boxes usually cover only a small volume. For this study, which requires a very large volume to provide enough statistics, we use the mock galaxies populated with the semi-analytical models (SAMs) for a dark-matter-only simulation with a large volume simulation box. Therefore, we choose the catalogues derived from the MultiDark Planck 2 simulation (MDPL2) \citep{1411.4001} by applying three different SAMs: \galacticus\ \citep{1008.1786}, \sag\ \citep{1801.03883}, and \sage\ \citep{1601.04709}, to it. These galaxy and halo catalogues are downloaded from the MultiDark database\footnote{https://www.cosmosim.org/} \citep{1109.0003}.

The MDPL2 simulation traces the evolution of $3840^3$ particles inside a cubic box with $1\hGpc$ on a side. The simulation follows the gravitational growth of these particles under $\Lambda$CDM model with the Planck 2016 cosmological parameters \citep{1502.01597}: $\Omega_m = 0.307$, $\Omega_b = 0.048$, $\Omega_{\Lambda} = 0.693$, $\sigma_8 = 0.823$, $n_s = 0.96$ and a dimensionless Hubble parameter $h = 0.678$ from $z=17$ to $z=0$ in 126 snapshots. 

Haloes, substructures and their corresponding tidal properties are then identified by the \rockstar\ phase-space temporal halo finder \citep{1110.4372}. This algorithm, an extension of the Friends-of-Friends algorithm in six phase-space dimensions, is currently the most successful in recovering all halo properties with a relatively high accuracy. Its unique ability to identify two spatially close haloes reduces the probability of missing an overdense area in this study. \rockstar\ has also been extensively tested to show its superior ability in recovering halo properties down to 20 particles comparing to other halo finders \citep{1104.0949}; this consistent accuracy is an important basis for creating model galaxy distributions.  Based on this halo catalogue, the merger tree is built using the {\sc consistent-tree} code \citep{1110.4372}. At $z=0$, we define clusters with $M_{\rm halo} > 10^{14} \hMsun$. There are 27,582 clusters in MDPL2. We refer to \citet{1411.4001} for the full halo mass function.

With the catalogue of haloes and their merging histories having been generated, three distinct SAMs (\galacticus, \sag, and \sage) are applied to provide separate catalogues of semi-analytical galaxies residing in those dark matter haloes. Galaxies are formed potentially within the branches of merger trees, and their corresponding properties are determined either by direct measurements or generated by a set of differential equations. The three SAMs share the same simulation data but generate different catalogues owing to their unique calibration settings and model designs, which all have their pros and cons. \galacticus\ \citep{1008.1786} has a powerful modular structure, and has advantages in its treatment of the SFR function and evolution. However, since it does not integrate the orbits of orphan galaxies but rather stores the positions where they are last found, its predicted clustering amplitudes are affected. This leads to a variation when selecting galaxies using mass or SFR thresholds. \sag, which originated from the Munich code \citep{SAG1} and was later developed by \cite{1801.03883}, has strengths in reproducing gas fractions and metallicity relations. Being calibrated by the Particle Swarm Optimisation technique \citep{1310.7034} and allowing gradual removal of the hot gas reservoir, \sag\ is able to achieve better agreement with observed atomic gas contents than most other SAMs. With the same origination as \sag, \sage\ \citep{1601.04709} has been updated and rebuilt from \cite{0508046}, calibrated using the stellar-mass function at $z=0$ and imposing two different sets of constraints, reproducing the stellar mass function accuracy into its strong point \citep[for a full comparison, see][]{Knebe2018}. The three SAMs share some similarities, such as a \cite{Chabrier2003} initial mass
function. They are also different in many aspects, such as the calibration and detailed model parameters, leading to their varied strengths: \sag\ can provide reasonable gas fractions and metallicity relations; \sage\ fits well the stellar mass function and stellar-to-halo mass relation; and \galacticus\
has its strength in the star formation rate function and evolution. It is worth pointing out that although two different dark matter halo mass definitions are used among three models, \cite{1505.04607} verified that there is little impact on the galaxy properties.

\subsection{Methods for protocluster searches} \label{method}
As structures in the universe form hierarchically with small objects merging to form larger objects, the most massive objects at $z=0$ -- galaxy clusters -- should form from an unusually large overdensity at high redshift, which defines a protocluster.
Thus, it is common to identify observed regions with high overdensity as protocluster candidates. As it is very hard to estimate the total density fluctuations in observation, different tracers are normally used to estimate the number or mass overdensity as an approximation of the total density fluctuation. As there are many different observables -- different galaxy properties, generally speaking -- that can be used as tracers, it is unclear which is a better choice. In this study, we focus on very high-redshift protoclusters, and thus we only follow \citetalias{jiang} to estimate the protocluster overdensities. 
First, the galaxy overdensity is defined as usual by 
\begin{equation}
    \delta_g = \frac{\rho - \Bar{\rho}}{\Bar{\rho}} = \frac{\rho}{\Bar{\rho}} - 1, 
\end{equation}
where $\rho$ is the galaxy number density within a certain region and $\Bar{\rho}$ is the average number density of all galaxies in the simulation box. 

Protoclusters are selected based on the overdensity of a fixed volume: $V = 35^3 \Mpc^3$ in this work.  We use a spherical volume instead of the cubical box as in \citetalias{jiang}, which is modified not only due to the isotropy of the universe, but also enables us to adopt a faster neighbour finding algorithm. In order to find giant protoclusters like the one detected by \citetalias{jiang}, we use the same search volume and overdensity $\delta_g = 5.6$. 

However, the set of protoclusters found in this way could have some overlap, since the test sphere may select out particles belonging to other distinct nearby galaxies duplicated in the area with large overdensity. We therefore exclude the potential overlap by removing less dense protoclusters with their member galaxies inside another more crowded protocluster from our final list. For example, if a member galaxy of two protoclusters having distance smaller than the sphere radius, i.e. $14.71\hMpc$, we will then compare the amount of galaxies in each region and only retain one belonging to the one with higher overdensity. 

The probability of finding a protocluster in 3D could thus be calculated by the proportion of the volume they occupy. 
\begin{equation}
    P_{\rm pc} = n_{\rm pc} \frac{V}{L^3}, \label{eq:ppc}
\end{equation}
where $n_{\rm pc}$ is the quantity of protoclusters being found, $L$ is the length of simulation box on one side, and $V$, as mentioned previously, is the volume we used to calculate the overdensity. For example, we have 4 haloes at $z=0$ from MDPL2 with a mass larger than the predicted cluster mass ($3.6 \times 10^{15} \Msun$) from \citetalias{jiang}. This 3D finding probability for the most massive clusters is $\sim$ 0.005 per cent.

As these SAM models are not calibrated at this high redshift, their mock galaxy properties may not match to observations. As shown in \citet{Knebe2018}, although the three models are generally in agreement with the observed cosmic star formation rate density, \sage\ is slightly higher while \sag\ is slightly lower than the observational data points at $z \simeq 6$. In \citetalias{jiang}, the overdensity is based on the Ly$\alpha$ flux-limited sample. The intrinsic Ly$\alpha$ luminosity can be converted to the galaxy SFR through the line emission ratio of Ly$\alpha$ to H$\alpha$ in Case B recombination and the relation between SFR and H$\alpha$ luminosity \citep[see ][for example]{Jiang2013}. Thus, it is easier to apply an SFR limit to select out Ly$\alpha$ galaxies from our mock sample. The SFR\;$> 1\Msunyr$ limitation is adopted in \citetalias{jiang} to match their Ly$\alpha$ flux-limited catalogue. However, it is worth noting that the approximate SFR values for the luminous LAEs may also be problematic \citep[for example][]{Pena-Guerrero2013}. Therefore, we supplement the original criterion in \citetalias{jiang} (listed as the first criterion below) for identifying protoclusters to include a further two criteria: 
\begin{itemize}
    \item[$\bullet$] Criterion 1: SFR $> 1\Msunyr$ (as originally proposed in \citetalias{jiang}). \\
    
    The SFR, corresponding to the Ly$\alpha$ emitters being used in observations \citep{Jiang2013}, provides us with a filter to categorise the mock galaxy catalogue. The same as \citetalias{jiang}, we use galaxies with SFR > $1\Msunyr$ as the Ly$\alpha$ emitters to find the potential protoclusters in the simulation. \cite{1310.2938} showed statistically that when an SFR cut is made at $1\Msunyr$, the probability of the overdense areas selected by $\delta_g = 5.6$ being a valid protocluster is 100\% up to $z=5$, which is much higher than when a cut is made on stellar mass. After discarding the overlapping galaxies, the protoclusters are then selected by overdensity $\delta_g \ge 5.6$. 
    \sepspace

    \item[$\bullet$] Criterion 2: $\Bar{\rho} = 3.48$. \\
    
    \citetalias{jiang} pointed out that the average galaxy number density $\Bar{\rho}$ at $z=5.7$ is 3.48 per ${35}^3 {\Mpc}^3$\footnote{This is estimated by 23 LAEs within a $35^3 \Mpc^3$ volume has an overdensity of 5.6 times the average galaxy number density.}. However, there is no guarantee that with SFR $> 1\Msunyr$, the mock galaxy catalogues can match the observed average galaxy number density. Actually, we find that $\Bar{\rho}$ is much higher than 3.48 with SFR $> 1\Msunyr$ for all three mock catalogues in Table~\ref{tab:table}. We therefore also vary the SFR threshold for each SAM catalogue such that the average density of the whole simulation box is equal to 3.48. We especially note here that the variance of SFR will not be directly proportional to changing the threshold for the observed Ly$\alpha$ emitters. This is because, for example, increasing the SFR will lead to more dust, which will affect the Ly$\alpha$ escape fraction \citep{Cai2014, Chiang2015}. However, as the intrinsic Ly$\alpha$ luminosity is empirically proportional to the galaxy SFR, we simply alter the SFR threshold to match the observed $\Bar{\rho}$ for this theoretical study. Using that SFR threshold, we further identify protoclusters with the same overdensity of $\delta_g \ge 5.6$. 
    \sepspace
    
    \item[$\bullet$] Criterion 3: a large protocluster sample with $n_{pc} \approx 3500$ 
    
    The predicted final mass of the protocluster from \citetalias{jiang} is $3.6 \times 10^{15} \Msun$, and there are only 4 clusters in MDPL2 with mass above this value. This is too small a sample for a statistical study (amounting to only about 0.015 per cent of all clusters at $z=0$). Thus we search for more protoclusters at $z=5.7$ to study the finding efficiency of the method. There are 3454 clusters in MDPL2 with $M_{z=0} > 10^{14.5} \hMsun$ which corresponds to $\sim$ 12.5 per cent of all MDPL2 clusters at $z=0$. By varying the SFR cut, we seek to identify a similar number of protoclusters at $z=5.7$, track them down to $z=0$ and study their final mass distributions. We note here that the same fixed volume is adopted for finding protocluster regions. This may affect protoclusters with smaller mass, but for simplicity we do not take this into account.
\end{itemize}

\section{Results} \label{Result}
In subsection \ref{subs:pc5.7}, we first present the basic information of the identified galaxy protoclusters at $z=5.7$ with the three approaches for each through the different SAM catalogues. In subsection \ref{subs:pc0}, these protoclusters are then traced down to $z=0$ to find their present-day properties and determine how many of them form massive clusters. 

\begin{table*}
    \centering
    \begin{tabular}{c|c|c|c|c|c|c|c|c|c|c|c}
        Criterion & SFR$^a$ [$\Msunyr$] & ${N_{g}}^b$ & ${\Bar{\rho}}^c$ & ${n_{\rm pc, 5.7}}^d$ & ${N_{\rm h, 5.7}}^e$ & ${P_{\rm pc, 5.7}\ \%}^f$ & ${n_{\rm pc, 0}}^g$ & ${N_{\rm h,0}}^h$ & ${N_{\rm C}}^i$ & ${N_{{\rm{MC}}}}^j$ & ${N_{{\rm{MMC}}}}^k$ \\
        \hline
        &&&&\galacticus\ &&&& \\
        SFR = 1 & 1.00 & 3,369,105 & 44.96 & 2 & 636 & 0.0027 & 2 & 18 & 2 & 2 & 0  \\
        $\Bar{\rho} = 3.48$ & 62.90 & 260,756 & 3.48 & 282 & 7379 & 0.38 & 252 & 1407 & 252 & 244 & 4 \\
        $n_{\rm pc} \approx 3500$ & 274.36  & 12,026 & 0.16 & 1283 & 2848 & 1.71 & 1269 & 2244 & 817 & 354 & 2 \\
        \hline
        &&&&\sag\ &&&& \\
        SFR = 1 & 1.00 & 2,192,869 & 29.26 & 19 & 3977 & 0.025 & 16 & 138 & 16 & 16 & 3 \\
        $\Bar{\rho} = 3.48$ & 5.26 & 260,756 & 3.48 & 588 & 16094 & 0.79 & 509 & 2756 & 509 & 488 & 4 \\
        $n_{\rm pc} \approx 3500$ & 23.65  & 22,705 & 0.30 & 4380 & 11644 & 5.85 & 4103 & 7003 & 3263 & 1382 & 4 \\
        \hline
        &&&&\sage\ &&&& \\
        SFR = 1 & 1.00 & 17,130,340 & 228.62 & 0 & 0 & 0.00 & 0 & 0 & 0 & 0 & 0 \\
        $\Bar{\rho} = 3.48$ & 48.48 & 260,756 & 3.48 & 445 & 11854 & 0.59 & 388 & 2214 & 388 & 372 & 4 \\
        $n_{\rm pc} \approx 35006$ & 115.08  & 22,705 & 0.30 & 4041 & 10195 & 5.39 & 3825 & 6970 & 2789 & 1131 & 4 \\
    \end{tabular}
    \caption{Results of three identification criteria on three SAM catalogues with \galacticus\ in the top block, \sag\ in the middle block, and \sage\ in the bottom block. The first column states each of the corresponding criterion being used. Note that the same overdensity ${\delta_g = 5.6}$ and search volume $V = 35^3 \Mpc^3$ as being used by \citetalias{jiang}.
    \newline ${}^a$ SFR thresholds under each situation with the first line in each block fixed to $1.00\Msunyr$ by the criterion. 
    \newline ${}^b$ Numbers of total galaxies in the whole simulation box with SFR larger than stated in the second column of each line at $z=5.7$. 
    \newline ${}^c$ The corresponding average number density of LAEs in a ${35}^3 {\Mpc}^3$ volume at $z=5.7$ with the second line of each block being fixed to 3.48 by the criterion. 
    \newline ${}^d$ Number of protoclusters regions being found at $z=5.7$. 
    \newline ${}^e$ Total number of distinct haloes in the protocluster regions at $z=5.7$.
    \newline ${}^f$ Probability of finding such a protocluster in 3D out of the whole simulation box. 
    \newline ${}^g$ The number of overdense regions at $z=0$ that are formed from these protoclusters. Note that we exclude these merged regions that is why this number is less-equal than column d.
    \newline ${}^h$ Total number of distinct haloes at $z=0$ evolved from the protocluster regions.
    \newline ${}^i$ Total number of clusters with $M \ge 10^{14} \hMsun$ from the regions of column g. Note that we only take the most massive halo in each region into account.
    \newline ${}^j$ Similar to column h, but for clusters with $M \ge 10^{14.5} \hMsun$.
    \newline ${}^k$ Similar to column h, but for clusters with $M \ge 3.6 \times 10^{15} \Msun$. 
    }
    \label{tab:table}
\end{table*}

\subsection{Identification of protoclusters at $z \simeq 5.7$} \label{subs:pc5.7}
The word `protocluster' is used differently by different authors, and no two individuals seem to agree on its exact definition.
But in general this word denotes a structure that will eventually collapse into a galaxy cluster with virialised mass greater than $10^{14} \Msun$ \citep{overdensemethod}. In practice, we use the galaxy number overdensity $\delta_g$ as a tracer to detect protoclusters. It is worth pointing out that using this tracer may cause some protoclusters to be missed since galaxies have not yet formed within a structure with density large enough compared to the environment. However, if a considerable quantity of protoclusters can still be detected under this strict criterion, then it shows a great potential to find further systems by other more subtle criteria. In MDPL2, there are 27,582 haloes at $z=0$ that can be classified as clusters by definition of $M > 10^{14} \hMsun$. In these clusters, there are 3454 haloes with $M > 10^{14.5} \hMsun$, 169 haloes with $M > 10^{15} \hMsun$ and 4 haloes with a mass larger than the predicted cluster mass ($3.6 \times 10^{15} \Msun$ from \citetalias{jiang}. Ideally, we expect this amount of protoclusters at higher redshift, at least these massive ones, to be detected. 

\subsubsection{Criterion 1: SFR = $1\Msunyr$}
\begin{figure*}
	\includegraphics[width=\linewidth]{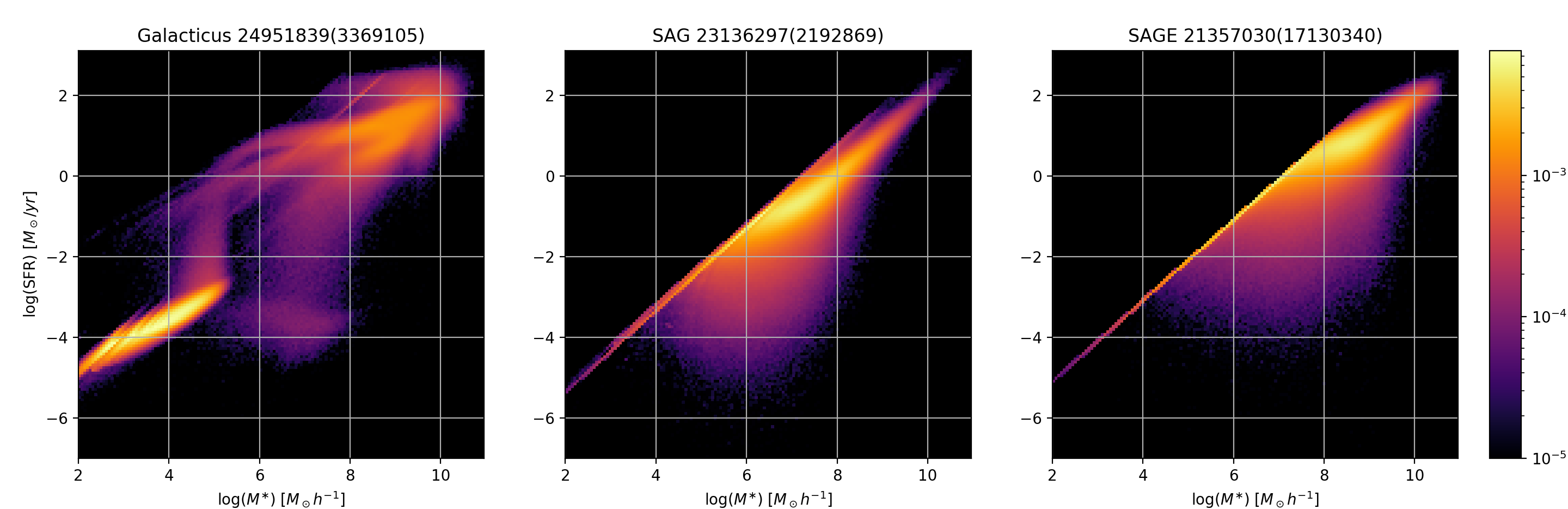}
    \caption{The distribution of galaxies at $z=5.7$ in three catalogues with the $y$-axis shows log(SFR) and the $x$-axis is log($\Mstar$). The colour shows the normalised density, which is presented in the colourbar to the right. Three panels are in the same scale and colour for easy comparison. The headline of each plot is the name and capacity of galaxies in it, where the number in the parenthesis is the number of galaxies with SFR larger than $1\Msunyr$. }
    \label{fig:figure1}
\end{figure*}

Through our settings, by imposing SFR = $1\Msunyr$, we find two protocluster candidates in the \galacticus\ catalogue, 19 in \sag, but none in \sage\ (see line 1 column e in each block in Table~\ref{tab:table}).

By focusing on column b, the total number of galaxies with SFR larger than $1\Msunyr$, we see comparable numbers for \galacticus\ and \sag\ catalogues but much larger (5-8 times) in \sage. The same difference is also imposed in column c -- the average galaxy number density -- in a different view. \cite{Knebe2018} predicted this phenomenon since they pointed out that although all three SAMs agree well with observational data in the area with $z$ from 0.6 to 2, \sage\ would over-predict the SFR density when $z > 4$ while \galacticus\ presents a good match through $z = 8.5$ and \sag\ shows a much lower SFR at high $z$. We further confirm this in Fig.~\ref{fig:figure1}, which shows the SFR distribution of galaxies in the three catalogues. A large proportion of galaxies in \sage\ have rather high SFR; while \sag\ has its galaxies distributed at the midpoint of SFR $\simeq 1 \Msunyr$; \galacticus\ shows two groups of galaxies at both very low SFR (around 0.0001 $\Msunyr$) and high SFR (around 10 $\Msunyr$). This explains well the comparison of the numbers of remaining galaxies in each catalogue with the SFR cut of $1\Msunyr$. We note here that although the galaxy stellar masses are shown as low as $100 \hMsun$, none of these low mass galaxies are selected for our analysis as the lowest SFR cut $1 \Msunyr$ implies a lowest stellar mass of $\sim 10^7 \hMsun$ for \sag\ and \sage\ or $\sim 10^5 \hMsun$ for \galacticus.

Due to the high average number density which means an even higher number density of LAEs in the protocluster region, it is not surprising to see that very few protoclusters are found in the three catalogues. As \sage\ has the highest average density, there is no protocluster found in this catalogue. Similarly, \citetalias{jiang} also reported that they are unable to find any protoclusters within their mock catalogue. The reason behind this should be an extremely high average overdensity, which requires an exaggerated overdensity of these protocluster regions. Given that only a few protoclusters (none for \sage) are identified with this criterion, the 3D finding probability $P_{\rm pc, 5.7}$ is extremely low as shown in column f. We further show the total number of distinct descendant haloes inside the protoclusters in column e. This number is larger than the total number of LAEs in the identified protocluster regions. It can give us a rough idea of how many satellite galaxies inside the resultant cluster, hence the mass of the cluster, assuming that they all fall into one cluster without merging. With over 200 massive subhaloes, we can expect these protoclusters will form massive clusters later.

\subsubsection{Criterion 2: $\Bar{\rho} = 3.48$}
\begin{figure*}
	\includegraphics[width=\linewidth]{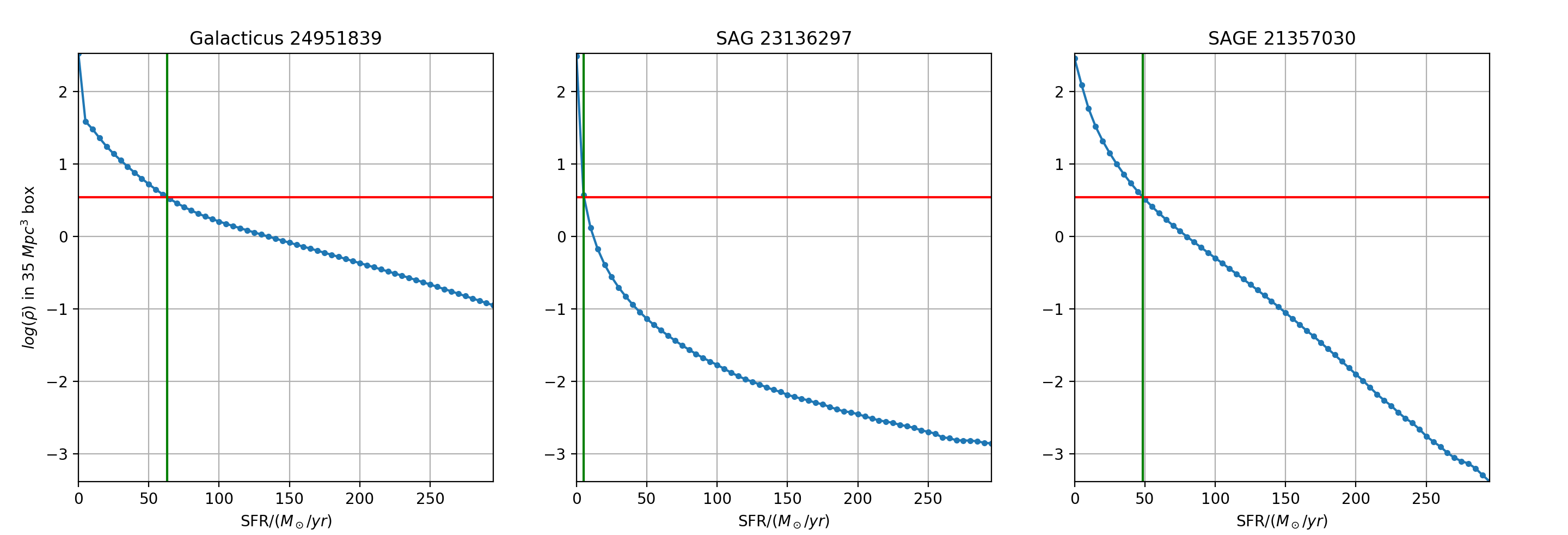}
    \caption{The changing of average galaxy number density $\Bar{\rho}$ as a function of SFR at $z=5.7$, where the red horizontal line denotes $\Bar{\rho} = 3.48$. Although a decreasing trend is presented in all three catalogues, the curves indicate a different dependence on the SFR threshold, which is intrinsically due to the treatment of galaxy formation in the three SAMs. It is clear that \galacticus\ has more high SFR galaxies than the other two models. }
    \label{fig:example_figure2}
\end{figure*}

Given the findings with Criterion 1, it is clear that the average number density $\Bar{\rho}$ is much higher than observation. Therefore, we restrict the average number density in the simulation box to be the same as the observed one by varying the SFR threshold -- Criterion 2.
With the requirement on the average galaxy number density to be 3.48, the numbers of LAEs remaining in the three catalogues become the same (column b), although the SFR thresholds for selecting them are quite different from each other. As stated in \cite{Knebe2018}, in the circumstance of low average galaxy density, despite the correlation becoming quite noisy, \sag, \sage, and \galacticus\ will have increased SFR thresholds to satisfy the average density condition, where this phenomenon is also found by us (see line 2 column a of each catalogue). Between these three models, \galacticus\ has the largest change of the SFR threshold, while \sag\ has the least change. This indicates that there are more galaxies with high SFR in \galacticus\ and \sage\ than \sag. A clear view of the effect of the SFR threshold on the average number density may be seen in Fig.~\ref{fig:example_figure2}. It is clear that \sag\ has the most significant drop with the SFR threshold increasing to around $10 \Msunyr$. 

Although the same $\Bar{\rho}$ and the same overdensity ${\delta_g = 5.6}$ is adopted in this criterion which guarantees the same number of LAEs in each protocluster region, the number of protoclusters from each catalogue can be different. This is because the SFR threshold is different, changing the clustering of the LAEs selected by the different SFR. With this low average number density, the requirement of the overdense regions, $\delta_g > 5.6$, can be relatively easy to achieve. Thus, more protoclusters are identified with this criterion: 282 protoclusters are found in \galacticus, 588 in \sag, and 445 in \sage\ (line 2 column e). However, with this criterion, we notice that the 3D probability of finding such protocluster regions is still very low. The total number of haloes (column e) inside the protocluster regions has a boost due to the increase of protoclusters. However, the average number of haloes in each protocluster is fixed to around 26 by definition. Thus, we expect a number of low mass clusters will be included in this sample.

\subsubsection{Criterion 3: a large protocluster sample with $n_{pc} \approx 3500$}
\begin{figure*}
	\includegraphics[width=\linewidth]{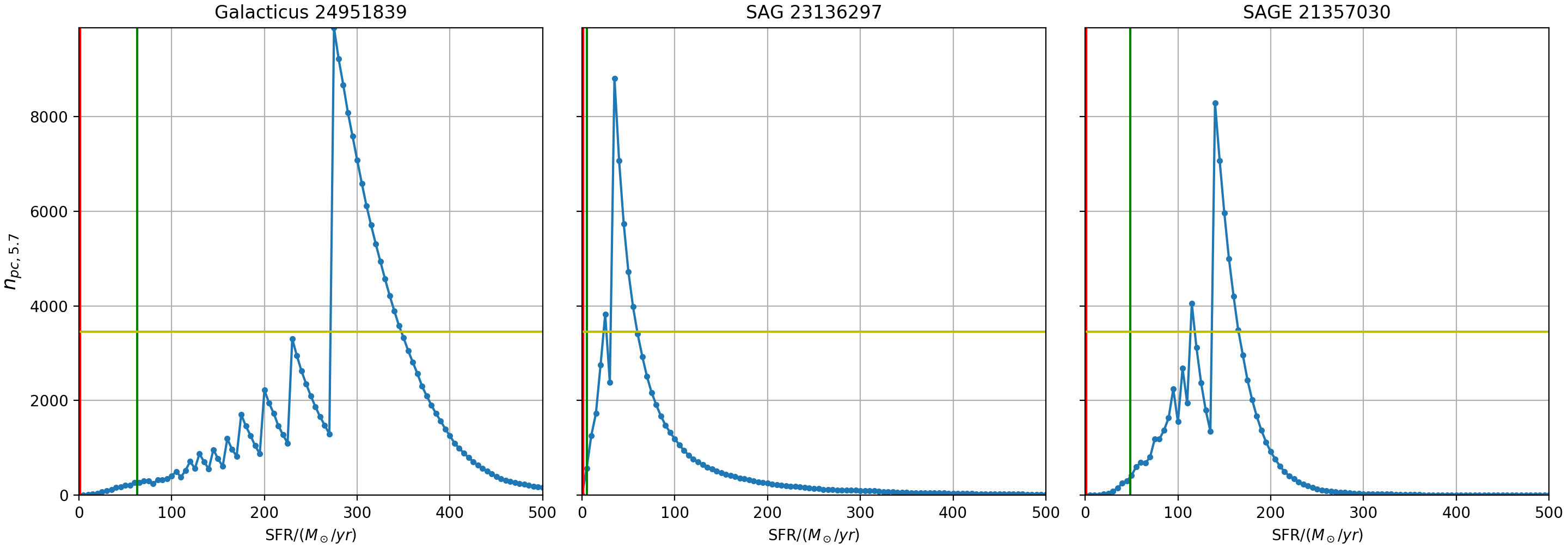}
    \caption{The effect of different SFR thresholds on the number of identified protoclusters at $z=5.7$. The horizontal yellow line shows the corresponding number  ($n_{pc} \approx 3500$) of protoclusters that yields the same number of clusters with $M_{z=0} \geq 10^{14.5} \hMsun$ in the simulation. As discussed in \Sec{method}, a fixed overdensity $\delta_g = 5.6$ is used to identify protocluster regions. The vertical red and green lines are the corresponding SFR threshold in Criterion 1 and Criterion 2. }
    \label{fig:example_figure3}
\end{figure*}

With criterion 3, we would like to identify more protoclusters for a statistical study of their final mass distribution. Therefore, we further vary the SFR threshold to produce a large protocluster sample with a higher 3D finding probability: $P_{\rm pc, 5.7} \approx 0.046$. Predicting from the figures of two previous criteria, we expect a much higher SFR threshold to reproduce this finding probability. However, as shown in Fig.~\ref{fig:example_figure3}, the number of protoclusters, and also the 3D finding probability, does not simply increase with the increased SFR threshold. It shows some zigzag features before its highest peak. Then, it decreases smoothly toward 0 as the SFR continually increases to several hundreds of $\Msunyr$. Since we keep the overdensity $\delta_g = 5.6$ to identify protoclusters, the zigzag increasing trend is probably caused by different reduction slopes in both the average number density and the density within the $35^3 \Mpc^3$ regions. The average number density dominates the increasing trend before the peak, while the density within the $35^3 \Mpc^3$ region dominates the decreasing trend after the peak. The expected number of protoclusters $n_{pc} = 3454$ is indicated by the horizontal yellow line, which crosses with the curve more than once. To simplify our study, we only use the first crossing point as the SFR threshold for selecting protoclusters. We note here that using the other crossing point presents similar results.

An even higher SFR threshold (column a), about 4 times greater than criterion 2, is needed for identifying such a large number of protoclusters. It is not surprising that the SFR thresholds differ between the three SAMs because of their different SFR distributions. With this criterion, 1283 protoclusters are found in \galacticus, 4380 in \sag, and 4041 in \sage\ (line 3 column e). We need to point out here that these numbers are not exactly the same as 3454, and thus the 3D finding probability is not exactly 4.6 per cent. This is because the curve is very sensitive to a tiny change of SFR threshold before the highest peak. We only use the SFR threshold that provides the closest finding probability to identify protoclusters without fine-tuning the SFR threshold for the exact fraction. However, we also note here that our results should be qualitatively unchanged. It is interesting to see that less or similar $N_{\rm halo, 5.7}$  (column e) compared to criterion 2 is shown. This indicates that criterion 3 may be not efficient in identifying massive clusters. 


\bigskip

We conclude here that the three SAMs give different numbers of protoclusters (or require different SFR thresholds when the number of protoclusters are set to be the same) when the same identifying criterion is applied. The detailed galaxy formation recipes from each model, especially the parameters that control the SFR, definitely play an important role in the number of identified protoclusters. \sag\ generally yields more protoclusters than the other two models, because it contains many fewer galaxies with high SFR ($\gsim 50 \Msunyr$). However, it is also clear that there are much larger differences between these criteria with which different observational results are matched. Nevertheless, none of the models can simultaneously match the observed average galaxy number density with the SFR threshold of $1 \Msunyr$.

\subsection{The protoclusters at present} \label{sec:z0} \label{subs:pc0}
\begin{figure*}
	\includegraphics[width=\linewidth]{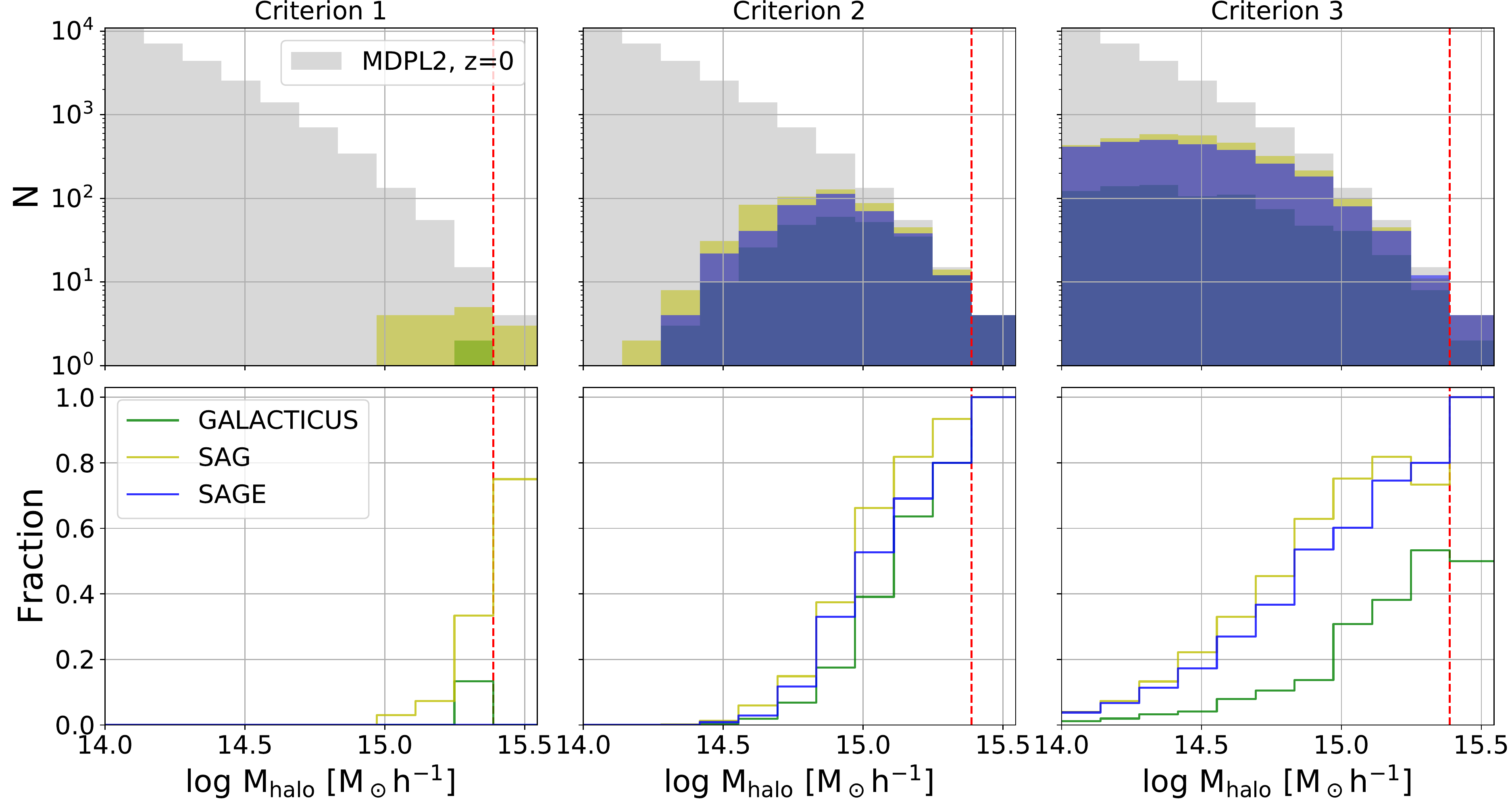}
    \caption{Upper: the histogram of the most massive haloes at $z=0$ resulted from all these protocluster regions at $z=5.7$ compared to the histogram of all the MDPL2 clusters at $z=0$; Lower: the number fraction in each mass bin of these identified haloes with respect to the MDPL2 clusters. The three criteria are shown in panels from left to right. The vertical red lines show the estimated final mass of the protocluster in \citetalias{jiang}. As indicated in the legends, different colours are for different SAMs with the gray histogram for the MDPL2 clusters.}
    \label{fig:prob_mmhmass}
\end{figure*}


An interesting question is whether the protoclusters at high redshift identified by applying this method to data would genuinely be the progenitors of massive galaxy clusters at $z=0$. Although we cannot address this question observationally, the merger tree from simulations allows us to provide a theoretical answer to this question. By matching the galaxies in the protocluster regions with their haloes in the MDPL2 halo catalogue, we track them down to $z=0$ and identify all their descendants using the merger tree provided by the query system in \url{https://www.cosmosim.org/}. We note here that some nearby protoclusters merged together at $z=0$. Therefore, fewer regions are presented in column g of Table~\ref{tab:table} compared to the number of protoclusters at $z=5.7$ (column d). There are significant numbers of haloes from $z=5.7$ (column e) in these protocluster regions, which merge to form massive clusters at $z=0$, leaving fewer distinct haloes (column h). We focus on the numbers of the most massive clusters in each protocluster region in the last three columns at different mass ranges: $\ge 10^{14} \hMsun, \ge 10^{14.5} \hMsun$ and $\ge 3.6\times 10^{15}\Msun$. It is not surprising to see fewer clusters at the massive end.

We further detail the distributions of the most massive haloes at $z=0$ from these protocluster regions in the upper panel of Fig.~\ref{fig:prob_mmhmass}. For comparison, we also include all the MDPL2 clusters. For criterion 1, both \galacticus\ and \sage\ fail to form such a massive cluster as claimed in \citetalias{jiang}. The final cluster masses from the two identified protoclusters in \galacticus\ are slightly smaller than the prediction in \citetalias{jiang}. Although \sag\ successfully identifies three out of the four most massive clusters in MDPL2, it also includes another 13 less massive clusters. Combining the two SAMs, we can confirm \citetalias{jiang}'s result but with a slightly lower cluster mass $2.82\pm0.92 \times 10^{15} \Msun$. For both criteria 1 and 2, the most massive descendant haloes in each of the protocluster regions have their halo masses greater than $10^{14} \hMsun$. However, there is a significant fraction ($\simeq 30$ per cent) of the most massive descendant haloes whose masses are less than $10^{14} \hMsun$ in Criterion 3. This means that the high SFR thresholds identify many regions of lower overdensity (due to the huge drop in the average number density). As there are more protoclusters identified from Criterion 1 to 3, more clusters from MDPL2 are identified at $z=0$. We note that these identified clusters seem to fill up the most massive end first. There seems to be little difference between the three SAMs on this final cluster mass distribution, except that no clusters are identified in \sage\ for criterion 1 and fewer clusters are found in \galacticus for criterion 3.

In the lower panel of Fig.~\ref{fig:prob_mmhmass}, we show the fraction of identified clusters in each mass bin. In agreement with the upper panel, the highest mass bin has the highest fraction with a significant drop as decreasing to lower mass bins with $\lsim 60$ per cent for $M_{\rm halo} < 10^{15} \hMsun$. With about 10 times more protoclusters identified in criterion 3 compared to criterion 2, however, the fraction for mass bins $10^{15}\hMsun \lsim M_{\rm halo} \lsim 10^{15.4}\hMsun$ is actually lower. This means that this method will not help to identify more massive cluster by simply including more protocluster regions through varying the SFR threshold. In addition to that, the detailed halo formation history may also play an important role. Actually, as presented in \cite{Wechsler2002} for example, the massive haloes are formed very recently, whereas their overdensity may not be that high for being picked up as a protocluster at such high redshift.

To further quantify the efficiency of this method for identifying massive clusters, we define two fractions: the purity,
\begin{equation}
    \mathbb{P} (M_s) = N_{rc}(M>M_s)/N_{pc},
\end{equation} 
which is the percentage of the protoclusters that result in cluster with mass $M > M_s$. Here, $N_{rc} (M>M_s)$ is the number of the cluster resulted from the protocluster with $M > M_s$ at $z=0$ with $M_s$ is the characterised mass and $N_{pc}$ is the total number of identified protoclusters; The completeness,
\begin{equation}
    \mathbb{C} (M_s) = N_{rc}(M>M_s)/N_{sc}(M>M_s),
\end{equation}
which is the ratio between $N_{rc}(M>M_s)$ and the total number of clusters $N_{sc}(M > M_s)$ in the simulation with $M > M_s$. Ideally, we expect the method has both high purity $\mathbb{P}$ and completeness $\mathbb{C}$ for a given characterised cluster mass. A high $\mathbb{C}$ means that most of the identified protoclusters will result in the expected cluster mass. This is useful for high quality protoclusters. While a high $\mathbb{C}$ will provide a more complete sample, i.e. $\mathbb{C} = 1$ means all the expected protoclusters (from $z=0$ clusters) are identified. For our case, with $M_s = 10^{15} \hMsun$, the purity $\mathbb{P} \simeq 0.938$ and the completeness $\mathbb{C} = 16/169 \simeq 0.089$ for the criterion 1 results from \sag. Identifying more protoclusters with criteria 2 and 3 gives a lower $\mathbb{P}$ (0.25 and 0.03, respectively) with a gain in $\mathbb{C}$ ($\sim$ 0.751). Furthermore, lowering $M_s$ will also increase $\mathbb{P}$ and decrease $\mathbb{C}$, which is easy to understand as this includes more clusters at $z=0$. These rates from different criteria and SAMs are detailed in Table~\ref{tab:table2}. We can see that no combination of model and criterion gives high $\mathbb{P}$ and completeness $\mathbb{C}$ simultaneously. An improvement in this overdensity method is needed.


\begin{table}
    \centering
    \begin{tabular}{c|c|c|c}
         $M_s$ & Criterion 1 & Criterion 2 & Criterion 3\\
         \hline
         & \galacticus\ & &\\
         $10^{15} \hMsun$ & 1.0 / 0.012 & 0.357 / 0.533 & 0.048 / 0.361\\
         $10^{14.5} \hMsun$ & 1.0 / 5.79 $\times 10^{-4}$ & 0.968 / 0.071 & 0.279 / 0.102\\
         $10^{14} \hMsun$ & 1 / 7.25 $\times 10^{-5}$ & 1.0 / 0.009 & 0.644 / 0.030 \\
         \hline
         & \sag\ & &\\
         $10^{15} \hMsun$ & 0.938 / 0.089 & 0.250 / 0.751 & 0.031 / 0.751 \\
         $10^{14.5} \hMsun$ & 1.0 / 0.005 & 0.959 / 0.141 & 0.337 / 0.400 \\
         $10^{14} \hMsun$ & 1.0 / 5.8 $\times 10^{-4}$ & 1.0 / 0.018 & 0.795 / 0.118 \\
         \hline
         & \sage\ & &\\
         $10^{15} \hMsun$ & - / - & 0.273 / 0.627 & 0.029 / 0.663 \\
         $10^{14.5} \hMsun$ & - / - & 0.959 / 0.108 & 0.296 / 0.327 \\
         $10^{14} \hMsun$ & - / - &  1.0 / 0.014 & 0.729 / 0.101 \\
    \end{tabular}
    \caption{The puritys $\mathbb{P}$ (the first value) with completenesss $\mathbb{C}$ (the second value) for different models and criteria at three characterised cluster masses.}
    \label{tab:table2}
\end{table}


\section{Discussion and Conclusions} \label{Conclusion}
In this paper, we use three different mock galaxy catalogues, namely \galacticus, \sag\ and \sage, from the MDPL2 galaxies presented in \cite{Knebe2018}. Following \citetalias{jiang}, we try to identify protocluster regions at $z=5.7$ with an overdensity $\delta_g = 5.6$. Similarly to \citetalias{jiang}, we also use the SFR threshold of $1\Msunyr$ (Criterion 1) for mimicking the LAE luminosity cut. In addition, we also consider two other criteria to identify the protoclusters by varying the SFR threshold: Criterion 2, a fixed average number density from \citetalias{jiang}; Criterion 3, a large sample of protoclusters $n_{pc} \approx 3500$. Benefiting from knowledge of the simulation merger tree, we trace these galaxies in the protocluster regions down to $z=0$ and study their final states. Our main findings can be summarised as follows: 
\begin{itemize}
    \item When the same identification criterion is applied, it is not surprising that the three models give different numbers of protoclusters. Qualitatively, the three SAMs are not very different from each other. As \sage\ has many more star forming galaxies (SFR $> 1 \Msunyr$) at $z=5.7$, which results in a very high average number density $\Bar{\rho}$, it fails to provide any protoclusters with Criterion 1. The other two models also give a very high $\Bar{\rho}$ compared to the observed result $\Bar{\rho} = 3.48$. Therefore, only countable protoclusters are identified. With the fixed $\Bar{\rho} = 3.48$ (criterion 2 with a higher SFR threshold), there are more identified protoclusters and even more with criterion 3 by design.
    \item Evolving these protoclusters from criterion 1 down to $z=0$, we find the final cluster mass is $2.82\pm0.92 \times 10^{15} \Msun$, which is consistent with the prediction from \citetalias{jiang} within $\pm 1 \sigma$. 
    \item The overdensity method applied with LAEs as the tracer is more suitable for identifying the most massive clusters. As shown in Fig.~\ref{fig:prob_mmhmass}, fewer clusters in the less massive bin are identified (resulting from the protoclusters) compared to the total number of clusters in the MDPL2 simulation, even with criterion 3 which has $\simeq 4000$ protoclusters. 
    \item To study the efficiency of this method, we defined two quantities: the purity $\mathbb{P}$ and the completeness $\mathbb{C}$ (see the text for details). For all the three SAMs and criteria, it is hard to have high $\mathbb{P}$ and $\mathbb{C}$ simultaneously.
\end{itemize}

As shown in this paper, the number of protoclusters identified in simulations depends on the type of SAM being used. Furthermore, none of the criteria give a high efficiency in identifying more protoclusters, which could due to the tracer selected in this paper or the method itself. We plan to work on an efficient method/tracer to identify the massive protoclusters in the next step. The most massive clusters and their surrounding environments selected from the MDPL2 simulation are re-simulated with different full physical baryonic models and are presented in the 300 project\footnote{\url{https://the300-project.org}} \citep{Cui2018,Wang2018,Mostoghiu2019,Arthur2019,Ansarifard2020,Haggar2020,Kuchner2020,Li2020,Knebe2020} for different studies. This provides us with the best catalogue for tracking these hydro-simulated clusters back to high redshift and identifying their progenitors.

\section*{Acknowledgements}
JQ is supported by the SoPA career development scholarship of the University of Edinburgh. WC \& JAP acknowledge support from the European Research Council under grant number 670193 (the COSFORM project). AK is supported by MICIU/FEDER under research grant PGC2018-094975-C21. He further acknowledges support from the Spanish Red Consolider MultiDark FPA2017-90566-REDC and further thanks Kings of Convenience for quiet is the new loud.

The CosmoSim database used in this paper is a service by the Leibniz-Institute for Astrophysics Potsdam (AIP).
The MultiDark database was developed in cooperation with the Spanish MultiDark Consolider Project CSD2009-00064. The authors gratefully acknowledge the Gauss Centre for Supercomputing e.V. (www.gauss-centre.eu) and the Partnership for Advanced Supercomputing in Europe (PRACE, www.prace-ri.eu) for funding the MultiDark simulation project by providing computing time on the GCS Supercomputer SuperMUC at Leibniz Supercomputing Centre (LRZ, www.lrz.de).
The Bolshoi simulations have been performed within the Bolshoi project of the University of California High-Performance AstroComputing Center (UC-HiPACC) and were run at the NASA Ames Research Center.

\section*{Data Availability}
The data underlying this article are available in the MultiDark database \url{https://www.cosmosim.org/}, at \url{https://doi.org/10.1002/asna.201211900}.




\bibliographystyle{mnras}
\bibliography{example} 

\begin{thebibliography}{}
\makeatletter
\relax
\def\mn@urlcharsother{\let\do\@makeother \do\$\do\&\do\#\do\^\do\_\do\%\do\~}
\def\mn@doi{\begingroup\mn@urlcharsother \@ifnextchar [ {\mn@doi@}
  {\mn@doi@[]}}
\def\mn@doi@[#1]#2{\def\@tempa{#1}\ifx\@tempa\@empty \href
  {http://dx.doi.org/#2} {doi:#2}\else \href {http://dx.doi.org/#2} {#1}\fi
  \endgroup}
\def\mn@eprint#1#2{\mn@eprint@#1:#2::\@nil}
\def\mn@eprint@arXiv#1{\href {http://arxiv.org/abs/#1} {{\tt arXiv:#1}}}
\def\mn@eprint@dblp#1{\href {http://dblp.uni-trier.de/rec/bibtex/#1.xml}
  {dblp:#1}}
\def\mn@eprint@#1:#2:#3:#4\@nil{\def\@tempa {#1}\def\@tempb {#2}\def\@tempc
  {#3}\ifx \@tempc \@empty \let \@tempc \@tempb \let \@tempb \@tempa \fi \ifx
  \@tempb \@empty \def\@tempb {arXiv}\fi \@ifundefined
  {mn@eprint@\@tempb}{\@tempb:\@tempc}{\expandafter \expandafter \csname
  mn@eprint@\@tempb\endcsname \expandafter{\@tempc}}}

\bibitem[\protect\citeauthoryear{{Andreon}, {Newman}, {Trinchieri}, {Raichoor},
  {Ellis}  \& {Treu}}{{Andreon} et~al.}{2014}]{cluster2-3}
{Andreon} S.,  {Newman} A.~B.,  {Trinchieri} G.,  {Raichoor} A.,  {Ellis}
  R.~S.,   {Treu} T.,  2014, \mn@doi [\aap] {10.1051/0004-6361/201323077},
  \href {https://ui.adsabs.harvard.edu/abs/2014A&A...565A.120A} {565, A120}

\bibitem[\protect\citeauthoryear{{Ansarifard} et~al.,}{{Ansarifard}
  et~al.}{2020}]{Ansarifard2020}
{Ansarifard} S.,  et~al., 2020, \mn@doi [\aap] {10.1051/0004-6361/201936742},
  \href {https://ui.adsabs.harvard.edu/abs/2020A&A...634A.113A} {634, A113}

\bibitem[\protect\citeauthoryear{{Arthur} et~al.,}{{Arthur}
  et~al.}{2019}]{Arthur2019}
{Arthur} J.,  et~al., 2019, \mn@doi [\mnras] {10.1093/mnras/stz212}, \href
  {https://ui.adsabs.harvard.edu/abs/2019MNRAS.484.3968A} {484, 3968}

\bibitem[\protect\citeauthoryear{Bah\'e, McCarthy, Balogh  \& Font}{Bah\'e
  et~al.}{2013}]{1210.8407}
Bah\'e Y.~M.,  McCarthy I.~G.,  Balogh M.~L.,   Font A.~S.,  2013, \mn@doi
  [Monthly Notices of the Royal Astronomical Society] {10.1093/mnras/stt109},
  430, 3017

\bibitem[\protect\citeauthoryear{Behroozi, Wechsler  \& Wu}{Behroozi
  et~al.}{2013}]{1110.4372}
Behroozi P.~S.,  Wechsler R.~H.,   Wu H.-Y.,  2013, \mn@doi [\apj]
  {10.1088/0004-637X/762/2/109}, 762, 109

\bibitem[\protect\citeauthoryear{Benson}{Benson}{2010}]{1008.1786}
Benson A.,  2010, \mn@doi [New Astronomy] {10.1016/j.newast.2011.07.004}, 17,
  175

\bibitem[\protect\citeauthoryear{B{u{a}}descu, Yang, Bertoldi, Zabludoff, Karim
   \& Magnelli}{B{u{a}}descu et~al.}{2017}]{1708.00447}
B{u{a}}descu T.,  Yang Y.,  Bertoldi F.,  Zabludoff A.,  Karim A.,   Magnelli
  B.,  2017, \mn@doi [The Astrophysical Journal] {10.3847/1538-4357/aa8220},
  845, 172

\bibitem[\protect\citeauthoryear{{Cai}, {Lapi}, {Bressan}, {De Zotti},
  {Negrello}  \& {Danese}}{{Cai} et~al.}{2014}]{Cai2014}
{Cai} Z.-Y.,  {Lapi} A.,  {Bressan} A.,  {De Zotti} G.,  {Negrello} M.,
  {Danese} L.,  2014, \mn@doi [\apj] {10.1088/0004-637X/785/1/65}, \href
  {https://ui.adsabs.harvard.edu/abs/2014ApJ...785...65C} {785, 65}

\bibitem[\protect\citeauthoryear{{Chabrier}}{{Chabrier}}{2003}]{Chabrier2003}
{Chabrier} G.,  2003, \mn@doi [\pasp] {10.1086/376392}, \href
  {https://ui.adsabs.harvard.edu/abs/2003PASP..115..763C} {115, 763}

\bibitem[\protect\citeauthoryear{{Chanchaiworawit} et~al.,}{{Chanchaiworawit}
  et~al.}{2017}]{Chanchaiworawit2017}
{Chanchaiworawit} K.,  et~al., 2017, \mn@doi [\mnras] {10.1093/mnras/stx782},
  \href {https://ui.adsabs.harvard.edu/abs/2017MNRAS.469.2646C} {469, 2646}

\bibitem[\protect\citeauthoryear{Chiang, Overzier  \& Gebhardt}{Chiang
  et~al.}{2013}]{1310.2938}
Chiang Y.-K.,  Overzier R.,   Gebhardt K.,  2013, \mn@doi [The Astrophysical
  Journal] {10.1088/0004-637x/779/2/127}, 779, 127

\bibitem[\protect\citeauthoryear{{Chiang} et~al.,}{{Chiang}
  et~al.}{2015}]{Chiang2015}
{Chiang} Y.-K.,  et~al., 2015, \mn@doi [\apj] {10.1088/0004-637X/808/1/37},
  \href {https://ui.adsabs.harvard.edu/abs/2015ApJ...808...37C} {808, 37}

\bibitem[\protect\citeauthoryear{Cora et~al.,}{Cora et~al.}{2018}]{1801.03883}
Cora S.~A.,  et~al., 2018, \mn@doi [Monthly Notices of the Royal Astronomical
  Society] {10.1093/mnras/sty1131}, 479, 2

\bibitem[\protect\citeauthoryear{Croton et~al.,}{Croton et~al.}{2006}]{0508046}
Croton D.~J.,  et~al., 2006, \mn@doi [Monthly Notices of the Royal Astronomical
  Society] {10.1111/j.1365-2966.2006.09994.x}, 367, 864

\bibitem[\protect\citeauthoryear{Croton et~al.,}{Croton
  et~al.}{2016}]{1601.04709}
Croton D.~J.,  et~al., 2016, \mn@doi [The Astrophysical Journal Supplement
  Series] {10.3847/0067-0049/222/2/22}, 222, 22

\bibitem[\protect\citeauthoryear{{Cucciati, O.} et~al.,}{{Cucciati, O.}
  et~al.}{2014}]{z3photometric}
{Cucciati, O.} et~al., 2014, \mn@doi [A\&A] {10.1051/0004-6361/201423811}, 570,
  A16

\bibitem[\protect\citeauthoryear{{Cui} et~al.,}{{Cui} et~al.}{2018}]{Cui2018}
{Cui} W.,  et~al., 2018, \mn@doi [\mnras] {10.1093/mnras/sty2111}, \href
  {https://ui.adsabs.harvard.edu/abs/2018MNRAS.480.2898C} {480, 2898}

\bibitem[\protect\citeauthoryear{Dey, Lee, Reddy, Cooper, Inami, Hong, Gonzalez
   \& Jannuzi}{Dey et~al.}{2016}]{1604.08627}
Dey A.,  Lee K.-S.,  Reddy N.,  Cooper M.,  Inami H.,  Hong S.,  Gonzalez
  A.~H.,   Jannuzi B.~T.,  2016, \mn@doi [The Astrophysical Journal]
  {10.3847/0004-637x/823/1/11}, 823, 11

\bibitem[\protect\citeauthoryear{Giacconi \& Rossi}{Giacconi \&
  Rossi}{1960}]{xraytheo}
Giacconi R.,  Rossi B.,  1960, \mn@doi [Journal of Geophysical Research
  (1896-1977)] {10.1029/JZ065i002p00773}, 65, 773

\bibitem[\protect\citeauthoryear{{Giacconi} et~al.,}{{Giacconi}
  et~al.}{1979}]{xrayexp}
{Giacconi} R.,  et~al., 1979, \mn@doi [apj] {10.1086/157110}, \href
  {https://ui.adsabs.harvard.edu/abs/1979ApJ...230..540G} {230, 540}

\bibitem[\protect\citeauthoryear{Gladders \& Yee}{Gladders \&
  Yee}{2000}]{redsequence}
Gladders M.~D.,  Yee H. K.~C.,  2000, \mn@doi [The Astronomical Journal]
  {10.1086/301557}, 120, 2148

\bibitem[\protect\citeauthoryear{Gobat et~al.,}{Gobat
  et~al.}{2011}]{cluster2-1}
Gobat R.,  et~al., 2011, \mn@doi [Astronomy and Astrophysics]
  {10.1051/0004-6361/201016084}, 526, A133

\bibitem[\protect\citeauthoryear{{Haggar}, {Gray}, {Pearce}, {Knebe}, {Cui},
  {Mostoghiu}  \& {Yepes}}{{Haggar} et~al.}{2020}]{Haggar2020}
{Haggar} R.,  {Gray} M.~E.,  {Pearce} F.~R.,  {Knebe} A.,  {Cui} W.,
  {Mostoghiu} R.,   {Yepes} G.,  2020, \mn@doi [\mnras]
  {10.1093/mnras/staa273}, \href
  {https://ui.adsabs.harvard.edu/abs/2020MNRAS.492.6074H} {492, 6074}

\bibitem[\protect\citeauthoryear{Haines et~al.,}{Haines et~al.}{2015}]{haines}
Haines C.~P.,  et~al., 2015, \mn@doi [The Astrophysical Journal]
  {10.1088/0004-637x/806/1/101}, 806, 101

\bibitem[\protect\citeauthoryear{Harikane et~al.,}{Harikane
  et~al.}{2019}]{1902.09555}
Harikane Y.,  et~al., 2019, \mn@doi [The Astrophysical Journal]
  {10.3847/1538-4357/ab2cd5}, 883, 142

\bibitem[\protect\citeauthoryear{{Higuchi} et~al.,}{{Higuchi}
  et~al.}{2019}]{Higuchi2019}
{Higuchi} R.,  et~al., 2019, \mn@doi [\apj] {10.3847/1538-4357/ab2192}, \href
  {https://ui.adsabs.harvard.edu/abs/2019ApJ...879...28H} {879, 28}

\bibitem[\protect\citeauthoryear{{Jiang} et~al.,}{{Jiang}
  et~al.}{2013}]{Jiang2013}
{Jiang} L.,  et~al., 2013, \mn@doi [\apj] {10.1088/0004-637X/772/2/99}, \href
  {https://ui.adsabs.harvard.edu/abs/2013ApJ...772...99J} {772, 99}

\bibitem[\protect\citeauthoryear{Jiang et~al.,}{Jiang et~al.}{2018}]{jiang}
Jiang L.,  et~al., 2018, \mn@doi [Nature Astronomy]
  {10.1038/s41550-018-0587-9}, 2, 962

\bibitem[\protect\citeauthoryear{Klypin, Yepes, Gottlöber, Prada  \&
  Heß}{Klypin et~al.}{2016}]{1411.4001}
Klypin A.,  Yepes G.,  Gottlöber S.,  Prada F.,   Heß S.,  2016, \mn@doi
  [Monthly Notices of the Royal Astronomical Society] {10.1093/mnras/stw248},
  457, 4340

\bibitem[\protect\citeauthoryear{Knebe et~al.,}{Knebe et~al.}{2011}]{1104.0949}
Knebe A.,  et~al., 2011, \mn@doi [Monthly Notices of the Royal Astronomical
  Society] {10.1111/j.1365-2966.2011.18858.x}, 415, 2293

\bibitem[\protect\citeauthoryear{Knebe et~al.,}{Knebe
  et~al.}{2015}]{1505.04607}
Knebe A.,  et~al., 2015, \mn@doi [Monthly Notices of the Royal Astronomical
  Society] {10.1093/mnras/stv1149}, 451, 4029

\bibitem[\protect\citeauthoryear{{Knebe} et~al.,}{{Knebe}
  et~al.}{2018}]{Knebe2018}
{Knebe} A.,  et~al., 2018, \mn@doi [\mnras] {10.1093/mnras/stx2662}, \href
  {https://ui.adsabs.harvard.edu/abs/2018MNRAS.474.5206K} {474, 5206}

\bibitem[\protect\citeauthoryear{{Knebe} et~al.,}{{Knebe}
  et~al.}{2020}]{Knebe2020}
{Knebe} A.,  et~al., 2020, \mn@doi [\mnras] {10.1093/mnras/staa1407}, \href
  {https://ui.adsabs.harvard.edu/abs/2020MNRAS.495.3002K} {495, 3002}

\bibitem[\protect\citeauthoryear{{Kravtsov} \& {Borgani}}{{Kravtsov} \&
  {Borgani}}{2012}]{Kravtsov2012}
{Kravtsov} A.~V.,  {Borgani} S.,  2012, \mn@doi [\araa]
  {10.1146/annurev-astro-081811-125502}, \href
  {https://ui.adsabs.harvard.edu/abs/2012ARA&A..50..353K} {50, 353}

\bibitem[\protect\citeauthoryear{{Kuchner} et~al.,}{{Kuchner}
  et~al.}{2020}]{Kuchner2020}
{Kuchner} U.,  et~al., 2020, \mn@doi [\mnras] {10.1093/mnras/staa1083}, \href
  {https://ui.adsabs.harvard.edu/abs/2020MNRAS.494.5473K} {494, 5473}

\bibitem[\protect\citeauthoryear{Kuiper, Venemans, Hatch, Miley  \&
  Röttgering}{Kuiper et~al.}{2012}]{1203.2196}
Kuiper E.,  Venemans B.~P.,  Hatch N.~A.,  Miley G.~K.,   Röttgering H. J.~A.,
   2012, \mn@doi [Monthly Notices of the Royal Astronomical Society]
  {10.1111/j.1365-2966.2012.20800.x}, 425, 801

\bibitem[\protect\citeauthoryear{Lee, Dey, Hong, Reddy, Wilson, Jannuzi, Inami
  \& Gonzalez}{Lee et~al.}{2014}]{1405.2620}
Lee K.-S.,  Dey A.,  Hong S.,  Reddy N.,  Wilson C.,  Jannuzi B.~T.,  Inami H.,
    Gonzalez A.~H.,  2014, \mn@doi [The Astrophysical Journal]
  {10.1088/0004-637x/796/2/126}, 796, 126

\bibitem[\protect\citeauthoryear{Li et~al.,}{Li et~al.}{2007}]{simprotoat6-3}
Li Y.,  et~al., 2007, \mn@doi [The Astrophysical Journal] {10.1086/519297},
  665, 187

\bibitem[\protect\citeauthoryear{{Li} et~al.,}{{Li} et~al.}{2020}]{Li2020}
{Li} Q.,  et~al., 2020, \mn@doi [\mnras] {10.1093/mnras/staa1385}, \href
  {https://ui.adsabs.harvard.edu/abs/2020MNRAS.495.2930L} {495, 2930}

\bibitem[\protect\citeauthoryear{Miller et~al.,}{Miller
  et~al.}{2018}]{1804.09231}
Miller T.~B.,  et~al., 2018, \mn@doi [Nature] {10.1038/s41586-018-0025-2}, 556,
  469

\bibitem[\protect\citeauthoryear{{Mostoghiu}, {Knebe}, {Cui}, {Pearce},
  {Yepes}, {Power}, {Dave}  \& {Arth}}{{Mostoghiu}
  et~al.}{2019}]{Mostoghiu2019}
{Mostoghiu} R.,  {Knebe} A.,  {Cui} W.,  {Pearce} F.~R.,  {Yepes} G.,  {Power}
  C.,  {Dave} R.,   {Arth} A.,  2019, \mn@doi [\mnras] {10.1093/mnras/sty3306},
  \href {https://ui.adsabs.harvard.edu/abs/2019MNRAS.483.3390M} {483, 3390}

\bibitem[\protect\citeauthoryear{Muldrew, Hatch  \& Cooke}{Muldrew
  et~al.}{2015}]{def}
Muldrew S.~I.,  Hatch N.~A.,   Cooke E.~A.,  2015, \mn@doi [Monthly Notices of
  the Royal Astronomical Society] {10.1093/mnras/stv1449}, 452, 2528

\bibitem[\protect\citeauthoryear{Ouchi et~al.,}{Ouchi
  et~al.}{2005}]{Ouchi_2005}
Ouchi M.,  et~al., 2005, \mn@doi [The Astrophysical Journal] {10.1086/428499},
  620, L1

\bibitem[\protect\citeauthoryear{Overzier}{Overzier}{2016}]{overdensemethod}
Overzier R.~A.,  2016, \mn@doi [The Astronomy and Astrophysics Review]
  {10.1007/s00159-016-0100-3}, 24, 14

\bibitem[\protect\citeauthoryear{Overzier, Guo, Kauffmann, De~Lucia, Bouwens
  \& Lemson}{Overzier et~al.}{2009}]{simprotoat6-2}
Overzier R.~A.,  Guo Q.,  Kauffmann G.,  De~Lucia G.,  Bouwens R.,   Lemson G.,
   2009, \mn@doi [Monthly Notices of the Royal Astronomical Society]
  {10.1111/j.1365-2966.2008.14264.x}, 394, 577

\bibitem[\protect\citeauthoryear{{Pe{\~n}a-Guerrero} \&
  {Leitherer}}{{Pe{\~n}a-Guerrero} \& {Leitherer}}{2013}]{Pena-Guerrero2013}
{Pe{\~n}a-Guerrero} M.~A.,  {Leitherer} C.,  2013, \mn@doi [\aj]
  {10.1088/0004-6256/146/6/158}, \href
  {https://ui.adsabs.harvard.edu/abs/2013AJ....146..158P} {146, 158}

\bibitem[\protect\citeauthoryear{{Planck Collaboration} et~al.,}{{Planck
  Collaboration} et~al.}{2016}]{1502.01597}
{Planck Collaboration} et~al., 2016, \mn@doi [A\&A]
  {10.1051/0004-6361/201525833}, 594, A24

\bibitem[\protect\citeauthoryear{Rhoads et~al.,}{Rhoads
  et~al.}{2004}]{narrowband}
Rhoads J.,  et~al., 2004, \mn@doi [The Astrophysical Journal] {10.1086/422094},
  611, 59

\bibitem[\protect\citeauthoryear{{Riebe} et~al.,}{{Riebe}
  et~al.}{2013}]{1109.0003}
{Riebe} K.,  et~al., 2013, \mn@doi [Astronomische Nachrichten]
  {10.1002/asna.201211900}, \href
  {https://ui.adsabs.harvard.edu/abs/2013AN....334..691R} {334, 691}

\bibitem[\protect\citeauthoryear{Ruiz et~al.,}{Ruiz et~al.}{2015}]{1310.7034}
Ruiz A.~N.,  et~al., 2015, \mn@doi [The Astrophysical Journal]
  {10.1088/0004-637x/801/2/139}, 801, 139

\bibitem[\protect\citeauthoryear{{Sembolini}, {De Petris}, {Yepes}, {Foschi},
  {Lamagna}  \& {Gottl{\"o}ber}}{{Sembolini} et~al.}{2014}]{Sembolini2014}
{Sembolini} F.,  {De Petris} M.,  {Yepes} G.,  {Foschi} E.,  {Lamagna} L.,
  {Gottl{\"o}ber} S.,  2014, \mn@doi [\mnras] {10.1093/mnras/stu554}, \href
  {https://ui.adsabs.harvard.edu/abs/2014MNRAS.440.3520S} {440, 3520}

\bibitem[\protect\citeauthoryear{Springel, White, Tormen  \&
  Kauffmann}{Springel et~al.}{2001}]{SAG1}
Springel V.,  White S. D.~M.,  Tormen G.,   Kauffmann G.,  2001, \mn@doi
  [Monthly Notices of the Royal Astronomical Society]
  {10.1046/j.1365-8711.2001.04912.x}, 328, 726

\bibitem[\protect\citeauthoryear{Springel et~al.,}{Springel
  et~al.}{2005}]{simprotoat6-1}
Springel V.,  et~al., 2005, \mn@doi [Nature] {10.1038/nature03597}, 435, 629

\bibitem[\protect\citeauthoryear{Stanford et~al.,}{Stanford
  et~al.}{2012}]{cluster2-2}
Stanford S.~A.,  et~al., 2012, \mn@doi [The Astrophysical Journal]
  {10.1088/0004-637x/753/2/164}, 753, 164

\bibitem[\protect\citeauthoryear{Steidel, Adelberger, Dickinson, Giavalisco,
  Pettini  \& Kellogg}{Steidel et~al.}{1998}]{steidel}
Steidel C.~C.,  Adelberger K.~L.,  Dickinson M.,  Giavalisco M.,  Pettini M.,
  Kellogg M.,  1998, \mn@doi [The Astrophysical Journal] {10.1086/305073}, 492,
  428

\bibitem[\protect\citeauthoryear{{Toshikawa} et~al.,}{{Toshikawa}
  et~al.}{2012}]{Toshikawa2012}
{Toshikawa} J.,  et~al., 2012, \mn@doi [\apj] {10.1088/0004-637X/750/2/137},
  \href {https://ui.adsabs.harvard.edu/abs/2012ApJ...750..137T} {750, 137}

\bibitem[\protect\citeauthoryear{{Toshikawa} et~al.,}{{Toshikawa}
  et~al.}{2014}]{Toshikawa2014}
{Toshikawa} J.,  et~al., 2014, \mn@doi [\apj] {10.1088/0004-637X/792/1/15},
  \href {https://ui.adsabs.harvard.edu/abs/2014ApJ...792...15T} {792, 15}

\bibitem[\protect\citeauthoryear{Toshikawa et~al.,}{Toshikawa
  et~al.}{2017}]{10.1093}
Toshikawa J.,  et~al., 2017, \mn@doi [Publications of the Astronomical Society
  of Japan] {10.1093/pasj/psx102}, 70

\bibitem[\protect\citeauthoryear{{Venemans} et~al.,}{{Venemans}
  et~al.}{2002}]{Venemans2002}
{Venemans} B.~P.,  et~al., 2002, \mn@doi [\apjl] {10.1086/340563}, \href
  {https://ui.adsabs.harvard.edu/abs/2002ApJ...569L..11V} {569, L11}

\bibitem[\protect\citeauthoryear{{Venemans} et~al.,}{{Venemans}
  et~al.}{2007}]{Venemans07}
{Venemans} B.~P.,  et~al., 2007, \mn@doi [A\&A] {10.1051/0004-6361:20053941},
  461, 823

\bibitem[\protect\citeauthoryear{{Wang} et~al.,}{{Wang}
  et~al.}{2018}]{Wang2018}
{Wang} Y.,  et~al., 2018, \mn@doi [\apj] {10.3847/1538-4357/aae52e}, \href
  {https://ui.adsabs.harvard.edu/abs/2018ApJ...868..130W} {868, 130}

\bibitem[\protect\citeauthoryear{{Wechsler}, {Bullock}, {Primack}, {Kravtsov}
  \& {Dekel}}{{Wechsler} et~al.}{2002}]{Wechsler2002}
{Wechsler} R.~H.,  {Bullock} J.~S.,  {Primack} J.~R.,  {Kravtsov} A.~V.,
  {Dekel} A.,  2002, \mn@doi [\apj] {10.1086/338765}, \href
  {https://ui.adsabs.harvard.edu/abs/2002ApJ...568...52W} {568, 52}

\makeatother
\end{thebibliography}








\bsp	
\label{lastpage}
\end{document}